\documentclass{article}

\usepackage[utf8]{inputenc}
\pdfoutput=1
\usepackage{amsmath} 
\usepackage{amssymb} 
\usepackage{amsthm} 
\usepackage{amsfonts} 
\usepackage{array} 
\usepackage{booktabs} 
\usepackage{braket} 
\usepackage{xcolor} 
\usepackage{cite} 
\usepackage{enumerate} 
\usepackage{extarrows} 
\usepackage{fancyhdr} 
\usepackage{float} 
\usepackage{geometry} 
\usepackage{graphicx} 
\usepackage{hyperref} 
\usepackage{listings} 
\usepackage{microtype} 
\usepackage{multirow} 
\usepackage{relsize} 
\usepackage{url} 

\usepackage[colorinlistoftodos]{todonotes}
\hypersetup{
	colorlinks=true,       
	linkcolor=blue,        
	citecolor=red,        
	filecolor=yellow,     
	urlcolor=blue         
}

\newgeometry{left=2.7cm, right=2.7cm, top=3.5cm, bottom=3.2cm}
\title{\textbf{Different Bootstrap Matrices in Many QM Systems}}
\date{} 

\begin{document}

\begin{titlepage}
\author{
	\Large Xihe Hu\footnote{huxihe@stumail.nwu.edu.cn}\\
	School of Physics, Northwest University, Xi'an 710127, China
}
\setcounter{footnote}{0}
\renewcommand{\thefootnote}{\arabic{footnote}}
\maketitle
\thispagestyle{empty}

\begin{abstract}
  The bootstrap is a technique recently developed to get energy eigenvalues of bound states and correlation functions. There are three crucial steps - recursive equations, positivity constraints, search space.
  We calculate recursive equations of many representative quantum mechanics systems, such as polynomial potential, exponential potential, Yukawa potential and electromagnetic potential.
  Two kinds of bootstrap matrices, which are about the coordinate and coupling of the coordinate with the momentum, and their ability of constraining equations are displayed.
  Nextly, we analyze possible questions in numerical search, including the importance of constraints and step length, eigen-energy level and the degeneracy of energy.
  Finally, we try to explain why the bootstrap work well by analyzing positivity constraints of creation operator and annihilation operator in harmonic oscillator.
  This article summarizes most knowledge of bootstrapping quantum mechanics (QM), and displays specific bootstrap equations and bootstrap matrices of different QM systems.
\end{abstract}
\end{titlepage}

\newpage
\tableofcontents

\newpage
\section{Introduction} 
The bootstrap are early applied to lattice theory \cite{anderson2017loop,lawrence2021bootstrapping,kazakov2022bootstrap}, conformal field \cite{poland2019conformal,guerrieri2021string} and matrix models \cite{lin2020bootstraps,kazakov2021analytic}.
Recently, the bootstrap is used to calculate energy eigenvalues of bound states and correlation functions in Quantum Mechanics (QM). Most papers about bootstrapping QM are inspired by Han \cite{han2020bootstrapping}.
Some papers calculate different systems \cite{aikawa2022application,du2022bootstrapping,berenstein2021bootstrapping1,berenstein2021bootstrapping2,bhattacharya2021numerical,han2020quantum,aikawa2022bootstrap}
and another considers different bootstrap matrices which are due to various trial operators\cite{aikawa2021comment}.

In the bootstrap, there are three crucial steps - recursive equations (bootstrap equations), positivity constraints and search space.
Firstly, one should write the form of physical quantity in specific research and then derive the bootstrap equations according to the nature of systems. Bootstrap equations are recursive form in QM.
It is the second step to acquire constraints of the system by mathematical and physical analysis. The constraints are positivity constraints in most conditions and bootstrap matrices can completely express positivity constraints.
Finally, one calculates bootstrap equations to get bootstrap matrices in search space and then excepts some space not satisfying constraints. It is for getting numerical solution that the last step is accomplished by computer.
The three steps are general process, so most papers about the bootstrap execute this process.

In this paper, we derive different bootstrap equations and get different bootstrap matrices in different QM systems by the three steps.

Firstly, we derive bootstrap equations of one and two variables in some QM systems having different potential energy. The Hamiltonian of QM is $H=p^2+V$.
In rectangular coordinates, we calculate three kinds of potential energy - polynomial potential, exponential potential and coupling potential of the both.
The polynomial potential energy, $V(x)=\sum_n a_n x^n$, includes $x^2$ (harmonic oscillator), $x^2+gx^4$ (quartic non-harmonic oscillator) and $g|x|^\nu$ (it can't be calculated by bootstrap).
The exponential potential energy $V(e^x)$ is $\sum_n a_n e^{nx}$, such as $V=g\text{cosh}(x)$ (non-relativistic Toda model).
The coupling of polynomial and exponential potential energy $V(x,e^x)$ is $\sum_{n,m}=x^n e^{mx}$.
In radial coordinates, we use the radial equation to make bootstrap equations same with them in rectangular coordinates, such as $-\frac{N}{r}$ (electrostatic force) and Yukawa potential $-a\frac{e^{-r}}{r}$.
In polar angle coordinates, because of the cyclical potential energy, trial operators must be cyclical. We choose correct trial operators and calculate trigonometric potential $V(\theta)=g\text{cos}x$.
We also get the bootstrap equations when there is electromagnetic field.
We find potential functions which can be calculated by bootstrap must be continuous and smooth functions (as same as having second order derivatives). For example, piecewise function $V(x)=g|x|^\nu$, the finite or infinite square well and the delta well can not be calculated by bootstrap.

Nextly, we analyze two kinds of bootstrap matrices $\mathcal{M}$ in every QM system. The simpler kind includes $\mathcal{M}(x^m,x^n)$, $\mathcal{M}(e^{mx},e^{nx})$, $\mathcal{M}(e^{mx},x^n)$ and $\mathcal{M}(e^{-im\theta},e^{in\theta})$.
Another includes $\mathcal{M}(x^m,p^n)$, $\mathcal{M}(e^{mx},p^n)$ and $\mathcal{M}(e^{im\theta},p^n)$. One finds that the simpler kind is more convenient to calculate numerical solution but another has stronger constraints.

Finally, we analyze some questions easily ignored in search space, such as the importance of constraints and step length, eigen-energy level and the degeneracy of energy.

Importantly, we try to explain why bootstrap work well. This part is directly inspired by Aikawa's paper \cite{aikawa2022bootstrap}. We find the positivity constraints in bootstrapping quantum mechanics as same as them in creation operator and annihilation operator, which can help us understand the reason of bootstrap working.

This paper summarizes most knowledge of bootstrapping QM and displays many useful results which mainly being bootstrap equations and bootstrap matrices.

\section{Bootstrapping Quantum Mechanics}
\subsection{Bootstrap Equations}
In quantum mechanics, the Hamiltonian is 
\begin{align}
  H=p^2+V
\end{align}
Because the Hamiltonian is hermitian for the wave functions, in the energy eigenstates, the commutator of the Hamiltonian $H$ and any operator $\mathcal{O}$ satisfy
\begin{gather}
  \Braket{H\mathcal{O}}=\Braket{\mathcal{O}H}=E\Braket{\mathcal{O}} \label{EO}\\
  \Braket{\left[H,\mathcal{O}\right]}=0 \label{HO0}
\end{gather}
When knowing the specific form of $\mathcal{O}$, we can derive bootstrap equations from \eqref{EO}\eqref{HO0}. Generally, bootstrap equations in QM are recursive equations, so if initial values including eigen-energy and the average of correlation functions are certain, other elements can be calculated by bootstrap.

\subsection{Bootstrap Matrices and Positivity Constraints}
\subsubsection{Bootstrap Matrix of One Operator}
Any operator $\mathcal{O}$ satisfies
\begin{align}
  \Braket{\mathcal{O}^\dagger \mathcal{O}} \ge 0 \label{OdO}
\end{align}
Generally, the operator $\mathcal{O}$ is a polynomial of one operator $A$.
\begin{align}
  \mathcal{O}=\sum_{m=0}^k a_m A^m \label{OamAm}
\end{align}
$\{a_m\}$ is any coefficient.
Bootstrap matrix is defined as 
\begin{align}
  \mathcal{M}_{mn}=\Braket{(A^\dagger)^m A^n} ,\quad m,n=0,1,2,...,k \label{MAmn}
\end{align}
which means that its size is $(k+1)\times(k+1)$. From \eqref{OdO}\eqref{OamAm}\eqref{MAmn}, we can find
\begin{align}
  \begin{aligned}
    \Braket{\mathcal{O}^\dagger \mathcal{O}} = \sum_{m,n=0}^k a_m^\ast \mathcal{M}_{mn} a_n = a^\dagger \mathcal{M} a \ge 0
  \end{aligned}
\end{align}
$a$ is a column vector. A unitary transformation $S$ turn bootstrap matrix $\mathcal{M}$ into diagonal matrix $\mathcal{M}'$.
\begin{gather}
  \begin{gathered}
    \mathcal{M}'=S\mathcal{M}S^\dagger ,\quad a'=Sa \\
    \Braket{\mathcal{O}^\dagger \mathcal{O}}={a'}^\dagger \mathcal{M}' a' = \sum_{m=0}^k {a'}_m^\ast {a'}_m M'_{mm} = \sum_{m=0}^k |{a'}_m|^2 M'_{mm} \ge 0
  \end{gathered}
\end{gather}
Because $\{a_m\}$ is arbitrary, all $\mathcal{M}'_{mm} $ must be positivity, $\mathcal{M}'_{mm} \ge 0$.

Bootstrap matrix of one operator is \eqref{MAmn}, and positivity constraint is
\begin{align}
  \forall \left( \mathcal{M} \right)_{\text{eigenvalue}} \ge 0 \label{Meigenvalue}
\end{align}
The larger $k$ of \eqref{MAmn} is, the stronger binding effect is. It should be emphasized that $\mathcal{M}_{mn}$ can be calculated by bootstrap equations.
Up to know, the positivity constraints are as if necessary and insufficient, but one can still gain good results by bootstrap. In the following section, we find positivity constraints may be necessary and sufficient.
\subsubsection{Bootstrap Matrix of Two Operators}
In another condition, the operator $\mathcal{O}$ is a coupling of operator $A$ and operator $B$.
\begin{align}
  \mathcal{O}=\sum_{m,n=0}^k a_m b_n A^m B^n
\end{align}
Because the commutator of $A$ and $B$ is not certain, $\mathcal{O}^\dagger \mathcal{O}$ is complex form.
\begin{align}
  \mathcal{O}^\dagger \mathcal{O} = \sum_{m_1,n_1,m_2,n_2=0}^k a_{m_1}^\ast b_{n_1}^\ast (B^\dagger)^{n_1} (A^\dagger)^{m_1} A^{m_2} B^{n_2} a_{m2} b_{n2}
\end{align}
For simplifying bootstrap matrix, we can define
\begin{gather}
  \begin{gathered}
    \mathcal{M}^1_{mn}=(B^\dagger)^n (A^\dagger)^m ,\quad \mathcal{M}^2_{mn}=A^m B^n ,\quad m,n=0,1,2,...,k  \\
    \mathcal{M}_{mn}=\Braket{\left( \mathcal{M}^1 \otimes \mathcal{M}^2 \right)_{mn}} ,\quad m,n=0,1,2,...,k^2
  \end{gathered} \label{M1mnM2mn}
\end{gather}
From \eqref{OdO}, we can find 
\begin{gather}
  \Braket{\mathcal{O}^\dagger \mathcal{O}}=\left(a \otimes b\right)^\dagger \mathcal{M} \left(a \otimes b\right) \ge 0
\end{gather}
So bootstrap matrix of two operators is \eqref{M1mnM2mn}, and positivity constraint is \eqref{Meigenvalue}.

Generally, bootstrap matrix of one operator is simpler, but when $k$ of two operators matrix as same as $k$ of one operator matrix, bootstrap matrix of two operators has stronger constraint and higher precision.

\subsection{Search Space}
This process should be completed by computer.
In search space, initial values are enumerated, and then other elements of bootstrap matrices can be calculated by bootstrap equations and initial values enumerated. We can get one numerical bootstrap matrix for a group of initial values.
Thus, many numerical bootstrap matrices include different group of initial numbers. Finally, we need to select matrices satisfying positivity constraints. These matrices's initial values include eigen-energy.
However, there are three questions about search space.

The first is the importance of constraints and search step length. In most systems, the search step length is more important than positivity constraints of bootstrap matrices.  
So when calculating numerical solves, we must improve the accuracy and control the constraints, which is equal to that the search step length must be very precise and $k$ of \eqref{MAmn} or \eqref{M1mnM2mn} is modest.
If the constraints are very strong ($k$ is very large) and the accuracy of enumerating initial values isn't very fine (the search step isn't precise), some eigen-energy may be ignored.

Secondly, how to determine the energy level is an important question. In search space, the results satisfying constraints constitute many closed loops. If the results are precise, every loop corresponds to one eigen-energy.
But one loop maybe includes two eigen-energy, so we only determine the order of these loops which have been calculated. When $k$ of \eqref{MAmn} or \eqref{M1mnM2mn} is enough large, there is one eigen-energy in one loop.
However, in this condition, the search step length must be enough accurate or else some eigen-energy may be ignored.

The last question is the degeneracy of eigen-energy. We think double degeneracy of eigen-energy which has eigenvalues $E_n$ and eigen-functions $f_{nl}$. Assuming a eigen-energy wave function $F_{n}$
\begin{gather}
  \begin{aligned}
    F_n=\sum_{l=l_1}^{l_2} a_l f_{nl} ,\quad \hat{H}f_{nl}=E_n f_{nl}
  \end{aligned}
\end{gather}
from \eqref{EO}, we can get
\begin{gather}
  \begin{align}
    \Braket{H\mathcal{O}}= E_n \sum_{l,l'=l_1}^{l_2} a_{l_1}^\ast a_{l_2} \Braket{f_{nl_1} |\mathcal{O}| f_{nl_2}}
  \end{align}
\end{gather}
It is equal to no degeneracy of eigen-energy. So the bootstrap is still applied to degenerate systems to calculate eigenvalues of energy. However, the average of correlation functions are not "real" but arbitrary.

\section{Polynomial Potential $V(x)$}
In QM systems having polynomial potential $V(x)$ of coordinates, the Hamiltonian and potential are
\begin{gather}
  H=p^2+V(x) \label{p2Vx}\\
  V(x)=\sum_l a_l x^l \nonumber
\end{gather}
\{$a_l$\} are arbitrary coefficients, but $V(x)$ must satisfy the system having bound states. So $V(x)=gx^\nu$ ($\nu$ is odd) should turn into 
\begin{align}
  V(x)=g\left|x\right|^\nu
\end{align}
However, $V(x)=g\left|x\right|^\nu$ can not be calculated by bootstrap. More details of bootstrap are displayed in the following.
\subsection{Bootstrap Equation and Matrix of One Variable $x$}
The derivation of this part's bootstrap equation is in Appendix \ref{xOne}. Bootstrap equation is 
\begin{align}
	\begin{aligned}
		4mE \Braket{x^{m-1}} - 4m \Braket{x^{m-1} V(x)} + m(m-1)(m-2) \Braket{x^{m-3}} - 2 \Braket{x^m V'(x)} =0
	\end{aligned}\label{xonebe}
\end{align}
We should notice normalization condition.
\begin{gather}
  \Braket{x^0}=1
\end{gather}
According to \eqref{MAmn}, we can structure bootstrap matrix of one hermitian operator $x$.
\begin{gather}
  \mathcal{M}_{mn}=\Braket{x^{m+n}} \quad m,n=0,1,2,...,k \label{xonebm}
\end{gather}
\eqref{xonebe} is a recursive equation, so except initial values, other $\Braket{x^{i+j}}$ can be calculated by \eqref{xonebe}.
And bootstrap matrix of $x$ satisfies constraint \eqref{Meigenvalue}.
\begin{gather}
  \forall \left(\mathcal{M}\right)_{\text{eigenvalue}} \ge 0
\end{gather}

\subsection{Bootstrap Equations and Matrix of Two Variables $x,p$}
The derivation of this part's bootstrap equations is in Appendix \ref{xTwo}. There are two bootstrap equations.
\begin{gather}
  E\Braket{x^m p^n} = -m(m-1)\Braket{x^{m-2}p^n} - 2im\Braket{x^{m-1}p^{n+1}} + \Braket{x^m p^{n+2}} + \Braket{V(x)x^m p^n} \label{xtwobe1} \\
  -m(m-1) \Braket{x^{m-2}p^n} - 2im \Braket{x^{m-1}p^{n+1}} + \Braket{x^m \left[V(x),p^n\right]} =0 \label{xtwobe2}
\end{gather}
When bootstrap matrix has two hermitian operators $x,p$, we don't use elimination method to simply bootstrap equations \eqref{xtwobe1}\eqref{xtwobe2}. According to \eqref{M1mnM2mn}, bootstrap matrix of $x,p$ is
\begin{gather}
  \begin{gathered}
    \mathcal{M}^1_{mn}=p^n x^m ,\quad \mathcal{M}^2_{mn}=x^m p^n ,\quad m,n=0,1,2,...,k  \\
    \mathcal{M}_{mn}=\Braket{\left( \mathcal{M}^1 \otimes \mathcal{M}^2 \right)_{mn}} ,\quad m,n=0,1,2,...,k^2
  \end{gathered} \label{xtwobm}
\end{gather}
\eqref{xtwobe1}\eqref{xtwobe2} are recursive equations of $x^mp^n$, but bootstrap matrix $\mathcal{M}$ includes $p^n x^m p^t$. So we need \eqref{pnxm} to calculate $p^n x^m p^t$.
\begin{gather}
  \begin{aligned}
    p^nx^mp^t=& \left[p^n,x^m\right]p^t + x^m p^{n+t}\\
    =& x^mp^{n+t} + \sum_{l=\text{min}\{0,n-m\}}^{n-1} \left(-i \right)^{n-l} \frac{n!m!}{l!(n-l)!(m-n+l)!} x^{m-n+l} p^{l+t}
  \end{aligned} \label{pnxmpt}
\end{gather}
From \eqref{xtwobe1}\eqref{xtwobe2}\eqref{pnxmpt}, we can get all elements of bootstrap matrix $\mathcal{M}$. Normalization condition:
\begin{gather}
  \Braket{x^0}=\Braket{p^0}=\Braket{x^0p^0}=1
\end{gather}
Bootstrap matrix of $x,p$ satisfies constraint \eqref{Meigenvalue}.
\begin{gather}
  \forall \left(\mathcal{M}\right)_{\text{eigenvalue}} \ge 0
\end{gather}

\subsection{Specific Systems}
\subsubsection{Harmonic Oscillator}
The research of bootstrapping harmonic oscillator is complete in Aikawa's paper \cite{aikawa2021comment}. The potential energy of harmonic oscillator is 
\begin{gather}
  V(x)=x^2
\end{gather}

According to \eqref{xonebe}, bootstrap equation of one various $x$ is
\begin{gather}
  m(m-1)(m-2)\Braket{x^{m-3}} + 4mE\Braket{x^{m-1}} - 4(m+1)\Braket{x^{m+1}}=0 \label{harosconebe}
\end{gather} 
Because the Hamiltonian of harmonic oscillator is even function, the wave function of eigen-energy is even. When $m$ is odd, the average values $\Braket{x^{m}}=0$. The useful information is in Table \ref{information}. Bootstrap matrix of $x$ is \eqref{xonebm} and positivity constraints are \eqref{Meigenvalue}.

From \eqref{xtwobe1}\eqref{xtwobe2} and Appendix \ref{harosc}, we can get bootstrap equations of two various $x,p$ are
\begin{gather}
  E\Braket{x^mp^n}=-m(m-1)\Braket{x^{m-2}p^n}-2im\Braket{x^{m-1}p^{n+1}}+\Braket{x^mp^{n+2}}+\Braket{x^{m+2}p^n}\label{harosctwobe1}\\
  -m(m-1)\Braket{x^{m-2}p^n}-2im\Braket{x^{m-1}p^{n+1}}+2in\Braket{x^{m+1}p^{n-1}}+n(n-1)\Braket{x^mp^{n-2}}=0 \label{harosctwobe2}
\end{gather}
The both of coordinates' wave function and momentum's wave function are even. So when $m+n$ is odd, $\Braket{x^mp^n}=0$. The information is in Table \ref{information}. Bootstrap matrix of $x,p$ is \eqref{xtwobm} and positivity constraints are \eqref{Meigenvalue}.
\subsubsection{Quartic Non-Harmonic Oscillator}
At the earliest, bootstrapping QM is studying quartic non-harmonic oscillator in Han's paper \cite{han2020bootstrapping}, but his paper only gives research method of one variable $x$ and we also supplement the method of two variables $x,p$.
 The potential energy of quartic non-harmonic oscillator is 
\begin{gather}
  V(x)=x^2+gx^4  
\end{gather}

According to \eqref{xonebe}, bootstrap equation of $x$ is
\begin{gather}
  m(m-1)(m-2)\Braket{x^{m-3}} + 4mE\Braket{x^{m-1}} - 4(m+1)\Braket{x^{m+1}} - 4g(m+2)\Braket{x^{m+3}} =0 \label{qnharosconebe}
\end{gather} 
The Hamiltonian of quartic non-harmonic oscillator is even function, so the wave function also is even. When $m$ is odd, the average values $\Braket{x^{m}}=0$. The information is in Table \ref{information}. Bootstrap matrix of $x$ is \eqref{xonebm} and positivity constraints are \eqref{Meigenvalue}.

From \eqref{xtwobe1}\eqref{xtwobe2} and Appendix \ref{qnharosc}, we can get bootstrap equations of two various $x,p$ are
\begin{gather}
  E\Braket{x^mp^n}=-m(m-1)\Braket{x^{m-2}p^n}-2im\Braket{x^{m-1}p^{n+1}}+\Braket{x^mp^{n+2}}+\Braket{x^{m+2}p^n}+g\Braket{x^{m+4}p^n} \label{qnharosctwobe1} \\
  \begin{aligned}
    &0= 4ign \Braket{x^{m+3}p^{n-1}} + 6gn(n-1) \Braket{x^{m+2}p^{n-2}} - 4ign(n-1)(n-2) \Braket{x^{m+1}p^{n-3}}\\
    &+ 2in \Braket{x^{m+1}p^{n-1}} + n(n-1) \Braket{x^m p^{n-2}} - gn(n-1)(n-2)(n-3) \Braket{x^mp^{n-4}}\\
    &- 2im \Braket{x^{m-1}p^{n+1}} - m(m-1) \Braket{x^{m-2}p^n}
  \end{aligned}\label{qnharosctwobe2}
\end{gather}
Because of same reason, the both of coordinates' wave function and momentum's wave function are even. So when $m+n$ is odd, $\Braket{x^mp^n}=0$. The information is in Table \ref{information}. Bootstrap matrix of $x,p$ is \eqref{xtwobm} and positivity constraints are \eqref{Meigenvalue}.
\subsubsection{$\boldsymbol{V(x)=g|x|^\nu}$ ($\boldsymbol{\nu \in}$ odd)}
The potential energy is 
\begin{gather}
  V(x)=g\left|x\right|^\nu, \quad \nu \text{ is odd}
\end{gather}

Firstly, we define $h_m$
\begin{align}
  h_{m-1}=\left\{
  \begin{array}{cl}
    0 & \text{if } m-1 = \text{odd},\\
    \displaystyle \int_0^{+\infty} 2x^{m+\nu-1} \,dx & \text{if } m-1 = \text{even}.
    \end{array} \right.
\end{align}
We can find $\Braket{x^{m-1}V(x)}$ and $\Braket{x^mV'(x)}$ are special elements and the parity of them are different with $\Braket{x^m}$ in \eqref{xonebe}.
\begin{gather*}
  \begin{aligned}
    \Braket{x^{m-1}V(x)}=\int_{-\infty}^{+\infty} x^{m-1}V(x) \,dx=\int_0^{+\infty} gx^{m+\nu-1} \,dx -\int_{-\infty}^0 gx^{m+\nu-1} \,dx= g h_{m-1}
  \end{aligned}\\
  \begin{aligned}
    \Braket{x^{m}V'(x)}=\int_{-\infty}^{+\infty} x^mV'(x) \,dx=g\nu\int_0^{+\infty} x^{m+\nu-1} \,dx -g\nu\int_{-\infty}^0 x^{m+\nu-1} \,dx =g\nu h_{m-1}
  \end{aligned}
\end{gather*}
In this condition, bootstrap equation is 
\begin{gather}
  4mE \Braket{x^{m-1}} + m(m-1)(m-2) \Braket{x^{m-3}} - 2g(2m-\nu)h_{m-1} =0 \label{simgrabe}
\end{gather}
This Hamiltonian is even function, so the wave function also is even. When $m$ is odd, the average values $\Braket{x^{m}}=h_m=0$. 
However, because of $h_m$, \eqref{simgrabe} is not closed and there are not constraints about $h_m$, so we can not get meaningful results. If potential function isn't continuous and smooth, extra and non-constrained initial numbers must be imported, which makes bootstrap equation not closed and meaningless, such as the finite square well, the infinite square well and the delta well.

So if one uses bootstrap, potential function must be continuous and smooth as same as have second order derivatives
\subsubsection{Radial Equation and Coulomb Potential}
The angular wave function $Y(\theta\phi)$ is same for spherically symmetric potential $V(r)$, so potential $V(r)$ only affects the radial wave function $R(r)$.
\begin{gather*}
  \frac{d}{dr}\left(r^2\frac{dR}{dr}\right) - r^2[V(r)-E]R=l(l+1)R \quad r\in(0,+\infty)
\end{gather*}
Defining
\begin{gather*}
  u(r)\equiv rR(r) \quad \text{ having } u(0)=u(\infty)=0 \text{ and } \int_0^{\infty} |u|^2 \,dr=1
\end{gather*}
the radial equation is
\begin{gather}
  -\frac{d^2u}{dr^2}+\left[\frac{l(l+1)}{r^2}+V\right]u=Eu
\end{gather}
So we can define radial momentum operator
\begin{gather}
  p_r=-i\frac{\partial }{\partial r}
\end{gather}
Radial momentum operator is a hermitian operator in radial coordinates and $\left[p_r,r\right]=-i$ is same with $\left[p_x,x\right]=-i$, which can be proved. In radial coordinates, the Hamiltonian and the effective potential are
\begin{gather}
  H=p_r^2+V_{\text{eff}}=p_r^2+\frac{l(l+1)}{r^2}+V(r) \quad r\in(0,+\infty)\\
  V_{\text{eff}}=\frac{l(l+1)}{r^2}+V(r)
\end{gather}
With radial equation, the form of the Hamiltonian and the commutator of radial momentum and radius are same with rectangular coordinates. What's more, the radial equation and $u(0)=u(\infty)=0$ ensure that \eqref{EO}\eqref{HO0} are true.

The effective potential of Coulomb potential is 
\begin{gather}
  V_{\text{eff}}=\frac{l(l+1)}{r^2}-\frac{N}{r}
\end{gather}
Berenstein's paper \cite{berenstein2021bootstrapping1} displays numerical results of Coulomb potential which are calculated by bootstrap of one variable $r$.

According to \eqref{xonebe}, Bootstrap equation of $r$ is 
\begin{gather}
  4mE\Braket{r^{m-1}}+2N(2m-1)\Braket{r^{m-2}}+(m-1)\left[m(m-2)-4l(l+1)\right]\Braket{r^{m-3}}=0 \label{Couonebe}
\end{gather}
The information is in Table \ref{information}. Bootstrap matrix of $r$ is \eqref{xonebm} and positivity constraints are \eqref{Meigenvalue}.

According to \eqref{xtwobe1}\eqref{xtwobe2} and Appendix \ref{coulomb}, bootstrap equations of $r,p_r$ are
\begin{gather}
  E\Braket{r^mp_r^n}=\left[l(l+1)-m(m-1)\right]\Braket{r^{m-2}p_r^n}-N\Braket{r^{m-1}p_r^n}-2im\Braket{r^{m-1}p_r^{n+1}}+\Braket{r^mp_r^{n+2}} \label{Coutwobe1}\\
  \begin{aligned}
    0=& -m(m-1) \Braket{r^{m-2}p_r^n} - 2im \Braket{r^{m-1}p_r^{n+1}} + N\sum_{t=0}^{n-1} i^t \frac{n!}{(n-t)!} \Braket{r^{m-1-t}p_r^{n-t}}\\
    & - l(l+1) \sum_{t=0}^{n-1} i^t \frac{n!}{(n-t)!(t+1)} \Braket{r^{m-2-t}p_r^{n-t}} 
  \end{aligned}\label{Coutwobe2}
\end{gather}
The information is in Table \ref{information}. Bootstrap matrix of $r,p_r$ is \eqref{xtwobm} and positivity constraints are \eqref{Meigenvalue}.

\section{Exponential Potential $V(e^x)$}
In QM systems having exponential potential $V(e^x)$, the Hamiltonian and potential are
\begin{gather}
  H=p^2+V(e^x) \\ 
  V(e^x)=\sum_l a_l e^{lx} \nonumber
\end{gather}
\{$a_l$\} are arbitrary coefficients and $V(e^x)$ is a polynomial of $e^x$, but $V(e^x)$ must satisfy system having bound states.
\subsection{Bootstrap Equation and Matrix of One Variable $e^x$}
The derivation of this part's bootstrap equation is Appendix \ref{exOne}. Bootstrap equation is 
\begin{align}
	\begin{aligned}
		(4mE+m^3)\Braket{e^{mx}} -4m\Braket{e^{mx}V(e^x)} -2\Braket{e^{mx}\frac{\partial V(e^x)}{\partial x}}=0
	\end{aligned}\label{exonebe}
\end{align}
Normalization condition:
\begin{gather}
  \Braket{e^{0x}}=1
\end{gather}
According to \eqref{MAmn}, we can structure bootstrap matrix of one hermitian operator $e^x$
\begin{gather}
  \mathcal{M}_{mn}=\Braket{e^{(m+n)x}} \quad m,n=0,1,2,...,k \label{exonebm}
\end{gather}
\eqref{exonebe} is a recursive equation, so all $\Braket{e^{(m+n)x}}$ can be calculated from \eqref{exonebe} and initial values.
And bootstrap matrix satisfies constraints \eqref{Meigenvalue}.
\begin{gather}
  \forall \left(\mathcal{M}\right)_{\text{eigenvalue}} \ge 0
\end{gather}

\subsection{Bootstrap Equations and Matrix of Two Variables $e^x,p$}
The derivation of this part's bootstrap equations is Appendix \ref{exTwo}. There are two bootstrap equations.
\begin{gather}
  E\Braket{e^{mx}p^n} = \Braket{e^{mx}p^{n+2}} - 2im\Braket{e^{mx}p^{n+1}} - m^2\Braket{e^{mx}p^n} + \Braket{V(e^x)e^{mx}p^n}  \label{extwobe1} \\
  -2im\Braket{e^{mx}p^{n+1}} - m^2\Braket{e^{mx}p^n} + \Braket{e^{mx}V(e^x)p^n} - \Braket{e^{mx}p^nV(e^x)} =0 \label{extwobe2}
\end{gather}
Depending \eqref{pneax}, we can calculate $\Braket{e^{mx}p^nV(e^x)}$.
As same as mentioned above, we don't simply bootstrap equations \eqref{extwobe1}\eqref{extwobe2}. According to \eqref{M1mnM2mn}, bootstrap matrix of $e^x,p$ is
\begin{gather}
  \begin{gathered}
    \mathcal{M}^1_{mn}=p^n e^{mx} ,\quad \mathcal{M}^2_{mn}=e^{mx} p^n ,\quad m,n=0,1,2,...,k  \\
    \mathcal{M}_{mn}=\Braket{\left( \mathcal{M}^1 \otimes \mathcal{M}^2 \right)_{mn}} ,\quad m,n=0,1,2,...,k^2
  \end{gathered} \label{extwobm}
\end{gather}
\eqref{extwobe1}\eqref{extwobe2} are recursive equations of $e^{mx}p^n$, but bootstrap matrix $\mathcal{M}$ includes $p^n e^{mx} p^t$. So we need \eqref{pneax} to calculate $p^n e^{mx} p^t$.
\begin{gather}
  \begin{aligned}
    p^ne^{mx}p^t=& e^{mx} \left(p-im\right)^n p^t = \sum_{l=0}^n \left(-im\right)^{n-l} \frac{n!}{l!(n-l)!} e^{mx}p^{l+t}
  \end{aligned} \label{pnemxpt}
\end{gather}
From \eqref{extwobe1}\eqref{extwobe2}\eqref{pnemxpt}, we can get all elements of bootstrap matrix $\mathcal{M}$. Normalization condition:
\begin{gather}
  \Braket{e^{0x}p^0}=1
\end{gather}
Bootstrap matrix of $e^x,p$ satisfies constraint \eqref{Meigenvalue}.
\begin{gather}
  \forall \left(\mathcal{M}\right)_{\text{eigenvalue}} \ge 0
\end{gather}

\subsection{Non-Relativistic Toda model}
Bootstrapping non-relativistic Toda model is shown by Du \cite{du2022bootstrapping}.
The potential energy of a non-relativistic Toda model is 
\begin{gather}
  V(e^x)=g\text{cosh}(x)=g\frac{e^x+e^{-x}}{2}
\end{gather}

According to \eqref{exonebe}, bootstrap equation of $e^x$ is
\begin{gather}
  -g(2m-1)\Braket{e^{(m-1)x}} +(4mE+m^3)\Braket{e^{mx}} -g(2m+1)\Braket{e^{(m+1)x}}=0 \label{Todaonebe}
\end{gather} 
The information is in Table \ref{information}. Bootstrap matrix of $e^x$ is \eqref{exonebm} and positivity constraints are \eqref{Meigenvalue}.

According to \eqref{extwobe1}\eqref{extwobe2} and Appendix \ref{Toda}, bootstrap equations are
\begin{gather}
  E\Braket{e^{mx}p^n} = \Braket{e^{mx}p^{n+2}} - 2im\Braket{e^{mx}p^{n+1}} - m^2\Braket{e^{mx}p^n} + \frac{g}{2}\Braket{e^{(m+1)x}p^n} + \frac{g}{2}\Braket{e^{(m-1)x}p^n} \label{Todatwobe1}\\
  \begin{aligned}
    0=& - 2im\Braket{e^{mx}p^{n+1}} - m^2\Braket{e^{mx}p^n} + \frac{g}{2}\Braket{e^{(m+1)x}p^n} + \frac{g}{2}\Braket{e^{(m-1)x}p^n}\\
      & - \frac{g}{2}\sum_{l=0}^n \frac{n!}{l!(n-l)!} \left(-i\right)^{n-l}\Braket{e^{(m+1)x}p^l} - \frac{g}{2}\sum_{l=0}^n \frac{n!}{l!(n-l)!} i^{n-l}\Braket{e^{(m-1)x}p^l}
  \end{aligned}\label{Todatwobe2}
\end{gather}
The information is in Table \ref{information}. Bootstrap matrix of $e^x,p$ is \eqref{extwobm} and positivity constraints are \eqref{Meigenvalue}.

\section{Coupling Potential $V(x,e^x)$}
For coupling potential $V(x,e^x)$, the Hamiltonian and potential are
\begin{gather}
  H=p^2+V(x,e^x) \\
  V(x,e^x)=\sum_{l_1, l_2} a_{l_1, l_2} x^{l_1} e^{l_2x} \nonumber
\end{gather}
\{$a_{l_1, l_2}$\} are arbitrary coefficients, but $V(x,e^x)$ must satisfy system having bound states.
\subsection{Bootstrap Equation and Matrix of Two Variables $x,e^x$}
The derivation of this part's bootstrap equation is Appendix \ref{xexTwo}. Bootstrap equation is 
\begin{align}
  \begin{aligned}
    0=& -(m^3+4mE)\Braket{e^{mx}x^n} - (3m^2n+4nE)\Braket{e^{mx}x^{n-1}} - 3mn(n-1)\Braket{e^{mx}x^{n-2}}\\
    & -n(n-1)(n-2)\Braket{e^{mx}x^{n-3}} + 4n\Braket{e^{mx}x^{n-1}V(x,e^x)} + 4m\Braket{e^{mx}x^nV(x,e^x)} \\
    & +2\Braket{e^{mx}x^n\frac{\partial V}{\partial x}}
  \end{aligned} \label{xextwobe}
\end{align}
Normalization condition:
\begin{gather}
  \Braket{x^0e^{0x}}=1
\end{gather}
From \eqref{MAmn}, we can structure bootstrap matrix of two hermitian operator $x,e^x$.
\begin{gather}
  \mathcal{M}_{mn}=\Braket{e^{mx}x^n} \quad m,n=0,1,2,...,k \label{xextwobm}
\end{gather}
\eqref{xextwobe} is a recursive equation, so all $\Braket{e^{mx}x^n}$ can be calculated from \eqref{xextwobe} and initial values.
Bootstrap matrix satisfies constraint \eqref{Meigenvalue}.
\begin{gather}
  \forall \left(\mathcal{M}\right)_{\text{eigenvalue}} \ge 0
\end{gather}

\subsection{Yukawa Potential}
The Yukawa Potential is
\begin{gather}
  V(r,e^r)=-a\frac{e^{-r}}{r}
\end{gather}
For radial equation, the effective potential
\begin{gather}
  V_\text{eff}= \frac{l(l+1)}{r^2} - a\frac{e^{-r}}{r}
\end{gather}

According to \eqref{xextwobe}, bootstrap equation of $r,e^r$ is
\begin{align}
  \begin{aligned}
    0=& -(m^3+4mE)\Braket{e^{mr}r^n} - (3m^2n+4nE)\Braket{e^{mr}r^{n-1}} + \left[ 4l(l+1)m - 3mn(n-1) \right]\Braket{e^{mr}r^{n-2}}\\
    & + \left[ 4l(l+1)(n-1) - n(n-1)(n-2) \right]\Braket{e^{mr}r^{n-3}} - 2a(2m-1)\Braket{e^{(m-1)r}r^{n-1}}\\
    & - 2a(2n-1)\Braket{e^{(m-1)r}r^{n-2}}
  \end{aligned}\label{Yukawatwobe}
\end{align}
However, we don't find suitable initial numbers to solves \eqref{Yukawatwobe}, which should be further studied.
The information is in Table \ref{information}. Bootstrap matrix of $r,e^r$ is \eqref{xextwobm} and positivity constraints are \eqref{Meigenvalue}.

\section{Polar Angle system $V(\theta)$}
In polar angle system, the Hamiltonian and cyclical potential are
\begin{gather}
  H=-\frac{\partial^2}{\partial \theta^2}+V(\theta)\\
  V(\theta)=V(\theta+2\pi)
\end{gather}
We can define polar momentum operator
\begin{gather}
  p_\theta=-i\frac{\partial}{\partial \theta}
\end{gather}
The polar momentum operator $p_\theta$ is hermitian operator in polar angle coordinates and $[p_\theta,\theta]=-i$ is same with $[p_x,x]=-i$.
Importantly, the polar momentum operator $p_\theta$ has similar physical meaning as angular momentum of z-axis $L_z=-i\frac{\partial}{\partial \phi}$. 
Rewriting the Hamiltonian:
\begin{gather}
  H=p_\theta^2+V(\theta)
\end{gather} 
If ensuring that \eqref{EO}\eqref{HO0} are true, the trial operator $\mathcal{O}$ must be cyclical, such as
\begin{gather}
  \mathcal{O}=e^{im\theta} \text{ or }e^{im\theta}p^n
\end{gather}
\subsection{Bootstrap Equation and Matrix of One Variable $e^{i\theta}$}
The derivation of this part's bootstrap equation is Appendix \ref{eiOne}. Bootstrap equation is 
\begin{align}
  (4mE-m^3) \Braket{e^{im\theta}} - 4m\Braket{e^{im\theta}V(\theta)} + 2i\Braket{e^{im\theta}\frac{\partial V}{\partial \theta}}=0 \label{eionebe}
\end{align}
Normalization condition:
\begin{gather}
  \Braket{e^{i0\theta}}=1
\end{gather}
Noting $e^{im\theta}$ is not hermitian, from \eqref{OdO}, we can structure bootstrap matrix of one operator $e^{i\theta}$ 
\begin{gather}
  \mathcal{M}_{mn}=\Braket{e^{i(-m+n)\theta}} \quad m,n=0,1,2,...,k \label{eionebm}
\end{gather}
\eqref{eionebe} is a recursive equation, so all $\Braket{e^{i(-m+n)\theta}}$ can be calculated from \eqref{eionebe} and initial values.
Bootstrap matrix satisfies constraint \eqref{Meigenvalue}.
\begin{gather}
  \forall \left(\mathcal{M}\right)_{\text{eigenvalue}} \ge 0
\end{gather}
Berenstein and Hulsey discuss one variable bootstrapping cyclical potential in \cite{berenstein2021bootstrapping2}

\subsection{Bootstrap Equation and Matrix of Two Variable $e^{i\theta},p_\theta$}
The derivation of this part's bootstrap equations is Appendix \ref{eiTwo}. There are two bootstrap equations.
\begin{gather}
  E\Braket{e^{im\theta}p_\theta^n} = \Braket{e^{im\theta}p_\theta^{n+2}} + 2m\Braket{e^{im\theta}p_\theta^{n+1}} + m^2\Braket{e^{im\theta}p_\theta^n} + \Braket{V(\theta)e^{im\theta}p_\theta^n}  \label{eitwobe1}\\
  2m\Braket{e^{im\theta}p_\theta^{n+1}} + m^2\Braket{e^{im\theta}p_\theta^n} + \Braket{e^{im\theta}V(\theta)p_\theta^n} - \Braket{e^{im\theta}p_\theta^nV(\theta)} =0 \label{eitwobe2}
\end{gather}
We calculate $\Braket{e^{im\theta}p_\theta^nV(\theta)}$ depending \eqref{pneax}.
As same as mentioned above, we don't simply bootstrap equations \eqref{eitwobe1}\eqref{eitwobe2}. According to \eqref{OdO}\eqref{M1mnM2mn}, bootstrap matrix of $e^{i\theta},p_\theta$ is
\begin{gather}
  \begin{gathered}
    \mathcal{M}^1_{mn}=p_\theta^n e^{-im\theta} ,\quad \mathcal{M}^2_{mn}=e^{im\theta} p_\theta^n ,\quad m,n=0,1,2,...,k  \\
    \mathcal{M}_{mn}=\Braket{\left( \mathcal{M}^1 \otimes \mathcal{M}^2 \right)_{mn}} ,\quad m,n=0,1,2,...,k^2
  \end{gathered} \label{eitwobm}
\end{gather}
\eqref{eitwobe1}\eqref{eitwobe2} are recursive equations of $e^{im\theta}p_\theta^n$, but bootstrap matrix $\mathcal{M}$ includes $p_\theta^n e^{imx} p_\theta^t$. So we need \eqref{pneax} to calculate $p_\theta^n e^{imx} p_\theta^t$.
\begin{gather}
  \begin{aligned}
    p_\theta^n e^{im\theta} p_\theta^t=& e^{im\theta} \left(p+m\right)^n p_\theta^t = \sum_{l=0}^n m^{n-l} \frac{n!}{l!(n-l)!} e^{im\theta}p_\theta^{l+t}
  \end{aligned} \label{pneimpt}
\end{gather}
From \eqref{eitwobe1}\eqref{eitwobe2}\eqref{pneimpt}, we can get all elements of bootstrap matrix $\mathcal{M}$. Normalization condition:
\begin{gather}
  \Braket{e^{i0\theta}p^0}=1
\end{gather}
Bootstrap matrix of $e^{i\theta},p_\theta$ satisfies constraint \eqref{Meigenvalue}.
\begin{gather}
  \forall \left(\mathcal{M}\right)_{\text{eigenvalue}} \ge 0
\end{gather}

\subsection{Trigonometric Potential $V(\theta)=gcos\theta$}
The trigonometric potential energy is
\begin{gather}
  V(\theta)=g\text{cos}(\theta)=g\frac{e^{i\theta}+e^{-i\theta}}{2}
\end{gather}

According to \eqref{eionebe}, bootstrap equation of $e^{i\theta}$ is
\begin{gather}
  -g(2m-1)\Braket{e^{i(m-1)\theta}} +(4mE-m^3)\Braket{e^{im\theta}} -g(2m+1)\Braket{e^{i(m+1)\theta}}=0 \label{cosonebe}
\end{gather} 
The information is in Table \ref{information}. Bootstrap matrix of $e^{i\theta}$ is \eqref{eionebm} and positivity constraints are \eqref{Meigenvalue}.

According to \eqref{eitwobe1}\eqref{eitwobe2} and Appendix \ref{cos}, bootstrap equations of $e^{i\theta},p_\theta$ are
\begin{gather}
  E\Braket{e^{im\theta}p_\theta^n} = \Braket{e^{im\theta}p_\theta^{n+2}} + 2m\Braket{e^{im\theta}p_\theta^{n+1}} + m^2\Braket{e^{im\theta}p_\theta^n} + \frac{g}{2}\Braket{e^{i(m+1)\theta}p_\theta^n} + \frac{g}{2}\Braket{e^{i(m-1)\theta}p_\theta^n} \label{costwobe1} \\
  \begin{aligned}
    0=& + 2m\Braket{e^{im\theta}p_\theta^{n+1}} + m^2\Braket{e^{im\theta}p_\theta^n} + \frac{g}{2}\Braket{e^{i(m+1)\theta}p_\theta^n} + \frac{g}{2}\Braket{e^{i(m-1)\theta}p_\theta^n} \\
      & - \frac{g}{2}\sum_{l=0}^n C_n^l \Braket{e^{i(m+1)\theta}p_\theta^l} - \frac{g}{2}\sum_{l=0}^n C_n^l \left(-i\right)^{n-l} \Braket{e^{i(m-1)\theta}p_\theta^l}
  \end{aligned}\label{costwobe2}
\end{gather}
The information is in Table \ref{information}. Bootstrap matrix of $e^{i\theta},p_\theta$ is \eqref{eitwobm} and positivity constraints are \eqref{Meigenvalue}.

\section{Electromagnetic Potential}
In classical electromagnetic system, a particle of charge $q$ uses canonical momentum $\mathbf{p'}$ to replace the momentum $\mathbf{p}$.
When there are electric and magnetic fields $\phi$ and $\mathbf{A}$, the canonical momentum and the Hamiltonian are
\begin{gather*}
  \mathbf{p}'=\mathbf{p}-q\mathbf{A}\\
  H=\frac{{p'}^2}{2m} + q\phi
\end{gather*}
In quantum system, one rewrite them.
\begin{gather}
  \mathbf{p}'= -i\hbar\nabla-q\mathbf{A}\\
  V=q\phi\\
  H= \left( \mathbf{p}-q\mathbf{A} \right)^2 + q\phi =  \left( -i\hbar\nabla-q\mathbf{A} \right)^2 + q\phi
\end{gather}

The bootstrap equations are rewrote.
\begin{gather}
  \Braket{[H,\mathcal{O}]} = \Braket{[p^2,\mathcal{O}]} - q\Braket{[\mathbf{A}\cdot\mathbf{p},\mathcal{O}]} - q\Braket{[\mathbf{p}\cdot\mathbf{A},\mathcal{O}]} + q^2\Braket{[A^2,\mathcal{O}]} + q\Braket{[\phi,\mathcal{O}]} = 0  \\
  \Braket{\mathcal{O}H} = \Braket{\mathcal{O}p^2} - q\Braket{\mathcal{O}\mathbf{A}\cdot\mathbf{p}} - q\Braket{\mathcal{O}\mathbf{p}\cdot\mathbf{A}} + q^2\Braket{\mathcal{O}A^2} + q\Braket{\mathcal{O}\phi} = E\Braket{\mathcal{O}}
\end{gather}
We only consider 1d space.When the electromagnetic field is a polynomial about $x$, we can make the trial operator $\mathcal{O}$ be $x^m$.
\begin{gather*}
  \Braket{\mathcal{O}H} = \Braket{x^mp^2} - 2q\Braket{x^mA(x)p} - q\Braket{x^m[p,A(x)]} + q^2\Braket{x^mA^2(x)} + q\Braket{x^m\phi(x)} = E\Braket{x^m}\\
  \Braket{[H,\mathcal{O}]} = -m(m-1)\Braket{x^{m-2}} - 2im\Braket{x^{m-1}p} - 2iqm\Braket{x^{m-1}A(x)} = 0
\end{gather*}
Nextly, the trial operator $\mathcal{O}$ is $x^mp$.
\begin{gather*}
  \begin{align*}
    \Braket{[H,\mathcal{O}]} =& -m(m-1)\Braket{x^{m-2}p} -2im\Braket{x^{m-1}p^2} +2iqm\Braket{x^{m-1}Ap} -2iq\Braket{x^{m}\frac{\partial A}{\partial x}p} -q\Braket{x^m\frac{\partial^2 A}{\partial x^2}}\\
    & +iq^2\Braket{x^m\frac{\partial A^2}{\partial x}} +iq\Braket{x^m\frac{\partial \phi}{\partial x}}=0
  \end{align*}
\end{gather*}
So we can get
\begin{gather}
  \begin{cases}
    \Braket{x^mp^2} - 2q\Braket{x^mAp} + iq\Braket{x^m\frac{\partial A}{\partial x}} + q^2\Braket{x^mA^2} + q\Braket{x^m} = E\Braket{x^m}\\
    2im\Braket{x^{m-1}p} = -m(m-1)\Braket{x^{m-2}} - 2iqm\Braket{x^{m-1}A}\\
    \begin{aligned}
      2im\Braket{x^{m-1}p^2} =& -m(m-1)\Braket{x^{m-2}p} +2iqm\Braket{x^{m-1}Ap} -2iq\Braket{x^{m}\frac{\partial A}{\partial x}p} -q\Braket{x^m\frac{\partial^2 A}{\partial x^2}}\\
      &+iq^2\Braket{x^m\frac{\partial A^2}{\partial x}} +iq\Braket{x^m\frac{\partial \phi}{\partial x}}
    \end{aligned}
  \end{cases}\label{EME}
\end{gather}
In electromagnetic field, the bootstrap equations are \eqref{EME}.

\section{The Same of Positivity Constraints between Ladder Operators and Bootstrapping QM}
Although knowing bootstrap can work well in quantum mechanics, one don't explain the reason. Luckily, in the paper \cite{aikawa2022bootstrap}, Aikawa etc. raise an effective idea that researching ladder operator can help us understand why bootstrap works well.
We will go to the same of positivity constraints between the both from creation and annihilation operators.

One usually write Hamiltonian with creation and annihilation operators. In Harmonic Oscillator, this form of Hamiltonian is
\begin{gather}
  H=\hbar \omega \left( a^\dagger a + \frac{1}{2} \right)
\end{gather} 
The commutator of $a^\dagger$ and $a$ is 
\begin{gather}
  \left[a,a^\dagger\right]=1 \label{comcreaanni}
\end{gather}
According to \eqref{comcreaanni}, that the interval of neighboring energy eigenvalues is $\hbar\omega$ can be found. 
\begin{gather}
  \begin{aligned}
    H a^\dagger\Ket{E_n}=\hbar\omega\left( n+1+\frac{1}{2} \right) a^\dagger\Ket{E_n}
  \end{aligned}\\
  \begin{aligned}
    H a\Ket{E_n}=\hbar\omega\left( n-1+\frac{1}{2} \right) a\Ket{E_n}
  \end{aligned}
\end{gather}
But up to now, one don't know which number $n$ is.

Nextly, it is the crucial step to calculate the minimum $n$ from positivity constraints.
\begin{gather}
  \Bra{E_n}a^\dagger a\Ket{E_n} \geqslant 0 \Rightarrow n \geqslant 0 \label{ada}
\end{gather}
So the energy eigenvalues are completely clear.

In bootstrapping quantum mechanics, the crucial point of deriving constraints is positivity constraints \eqref{OdO}.
The trial operator $\mathcal{O}$ is expressed with the coordinate $x$ and the momentum $p$ in the above, such as \eqref{xonebm},\eqref{xtwobm},\eqref{exonebm},\eqref{extwobm}.
But as we all know, the coordinates and the momentums can be expressed with the creation operator $a^\dagger$ and annihilation operator $a$.
When writing trial operator $\mathcal{O}$ with $a^\dagger$ and $a$, we can find the positivity constraints in bootstrapping QM \eqref{OdO} contain them in QM\eqref{ada}.
\begin{gather}
  \begin{aligned}
    \Braket{\mathcal{O}^\dagger\mathcal{O}} \geqslant 0 \overset{\text{contain}}{\rightarrow} \Braket{a^\dagger a} \geqslant 0
  \end{aligned}
\end{gather}
We can find positivity constraints may be necessary and sufficient.In the abstract, one can calculate not only number solution but also exact solution in bootstrapping QM and clearly understand why bootstrap works.
Aikawa's paper \cite{aikawa2022bootstrap} shows how to solve exact results with bootstrap in harmonic oscillator, which gives a convincing conclusion - at least, the positivity constraints are necessary and sufficient in harmonic oscillator.
So one can better understand why bootstrap works well by the same of positivity constraints between the both and exact values in harmonic oscillator.

\section{Conclusion}
In this paper, we mainly derive different bootstrap equations, analyze two major kinds of bootstrap matrices in many quantum mechanics systems and explain why bootstrap works well.
Most QM systems which can be calculate by bootstrap are displayed. In other systems, why not using bootstrap may be that bootstrap equations are not closed, that initial numbers are not discovered or that constraints are not suitable.
When not knowing wave function, one can get eigen-energy and the average of correlation functions by bootstrap, which is surprising.
Anyway, the bootstrap is a good technique and has huge prospects.

\begin{table}[!]
  \centering
  \renewcommand\arraystretch{1.5}
  \caption{The information of bootstrap equation}
  \label{information}
  \begin{tabular}{c|c|c|l|l}
    \toprule
    Potential & Equations & Initial Elements & Known Elements & Elements Calculated\\
    \midrule
    \multirow{4}{*}{$x^2$} & \multirow{2}{*}{\eqref{harosconebe}} & \multirow{4}{*}{$E$} & $\Braket{x^0}=1$, & $\Braket{x^2}$, $\Braket{x^4}$, $\Braket{x^6}$,...,\\
     & & & $\Braket{x^m}=0$ ($m\in$ odd) &  $\Braket{x^k}$, ($k\in$ even)\\
     \cmidrule(r){2-2} \cmidrule(r){4-5}
    & \multirow{2}{*}{\eqref{harosctwobe1}\eqref{harosctwobe2}} & & $\Braket{x^0}=\Braket{p^0}=\Braket{x^0p^0}=1$ & \multirow{2}{*}{$\Braket{x^mp^n}$, ($m+n\in$ even)}\\
    & & & $\Braket{x^mp^n}=0$ ($m+n\in$ odd) &  \\
    \midrule
    \multirow{4}{*}{$x^2+gx^4$} & \multirow{2}{*}{\eqref{qnharosconebe}} & \multirow{4}{*}{$E,\Braket{x^2}$} & $\Braket{x^0}=1$, & $\Braket{x^4}$, $\Braket{x^6}$, $\Braket{x^8}$,...,\\
    & & & $\Braket{x^m}=0$ ($m\in$ odd) &  $\Braket{x^k}$, ($k\in$ even)\\
    \cmidrule(r){2-2} \cmidrule(r){4-5}
    & \multirow{2}{*}{\eqref{qnharosctwobe1}\eqref{qnharosctwobe2}} & & $\Braket{x^0}=\Braket{p^0}=\Braket{x^0p^0}=1$ & \multirow{2}{*}{$\Braket{x^mp^n}$, ($m+n\in$ even)}\\
    & & & $\Braket{x^mp^n}=0$ ($m+n\in$ odd) &  \\
    \midrule
    & \multirow{2}{*}{\eqref{Couonebe}} & \multirow{4}{*}{$E$} &  \multirow{2}{*}{$\Braket{r^0}=1$} & $\Braket{r^{-1}}$, $\Braket{r}$, $\Braket{r^2}$,\\
    $-\frac{N}{r}$ & & & &  $\Braket{r^3}$,..., $\Braket{r^k}$\\
    \cmidrule(r){2-2} \cmidrule(r){4-5}
    $V_\text{eff}=\frac{l(l+1)}{r^2}-\frac{N}{r}$ & \multirow{2}{*}{\eqref{Coutwobe1}\eqref{Coutwobe2}} & & \multirow{2}{*}{$\Braket{r^0}=\Braket{p_r^0}=\Braket{r^0p_r^0}=1$} & \multirow{2}{*}{$\Braket{r^mp_r^n}$}\\
    & & &  &  \\
    \midrule
    \multirow{4}{*}{$g\text{cosh}(x)$} & \multirow{2}{*}{\eqref{Todaonebe}} & \multirow{4}{*}{$E,\Braket{e^x}$} & \multirow{2}{*}{$\Braket{e^{0x}}=1$} & $\Braket{e^{2x}}$, $\Braket{e^{3x}}$, $\Braket{e^{4x}}$,...,\\
    & & & &  $\Braket{e^{kx}}$\\
    \cmidrule(r){2-2} \cmidrule(r){4-5}
    & \multirow{2}{*}{\eqref{Todatwobe1}\eqref{Todatwobe2}} & &  \multirow{2}{*}{$\Braket{e^{0x}}=\Braket{p^0}=\Braket{e^{0x}p^0}=1$} & \multirow{2}{*}{$\Braket{e^{mx}p^n}$}\\
    & & & &  \\
    \midrule
    $-a\frac{e^{-r}}{r}$ & \multirow{2}{*}{\eqref{Yukawatwobe}} & \multirow{2}{*}{Unknown} &  \multirow{2}{*}{$\Braket{r^0}=\Braket{e^{0r}}=\Braket{e^{0r}r^0}=1$} & \multirow{2}{*}{$\Braket{e^{mr}r^n}$}\\
    $V_\text{eff}=\frac{l(l+1)}{r^2}-a\frac{e^{-r}}{r}$ & & & & \\
    \midrule
    \multirow{4}{*}{$g\text{cos}(\theta)$} & \multirow{2}{*}{\eqref{cosonebe}} & \multirow{4}{*}{$E,\Braket{e^{i\theta}}$} & \multirow{2}{*}{$\Braket{e^{i0\theta}}=1$} & $\Braket{e^{-ik\theta}}$, $\Braket{e^{-i(k-1)\theta}}$,...,\\
    & & & & $\Braket{e^{i(k-1)\theta}}$, $\Braket{e^{ik\theta}}$\\
    \cmidrule(r){2-2} \cmidrule(r){4-5}
    & \multirow{2}{*}{\eqref{costwobe1}\eqref{costwobe2}} & &  \multirow{2}{*}{$\Braket{e^{i0\theta}}=\Braket{p_\theta^0}=\Braket{e^{i0\theta}p_\theta^0}=1$} & \multirow{2}{*}{$\Braket{e^{-im\theta}p_\theta^n}$, $\Braket{e^{im\theta}p_\theta^n}$}\\
    & & & &  \\
    \bottomrule
  \end{tabular}
\end{table}

\newpage
\onecolumn
\newpage
\appendix

\section*{Appendix}
\addcontentsline{toc}{section}{Appendices}
\renewcommand{\thesubsection}{\Alph{subsection}}

\subsection{The Fundamental Knowledge}\label{TFK}
Firstly, we calculate the commutator relation of momentum $p^n$ and coordinate $x^n$.
\begin{align}
  &\begin{aligned}
      [p^n,x^m]=&\left(-i\right)^n \left[  \frac{\partial^n}{\partial x^n} (x^m f) - x^m \frac{\partial^n f}{\partial x^n}  \right]\\
      =&\left(-i \right)^n \left[  \sum_{l=0}^n C_n^l \left(\frac{\partial^{n-l}}{\partial x^{n-l}}x^m\right) \frac{\partial^l f}{\partial x^l} - x^m \frac{\partial^n f}{\partial x^n}  \right] \\
      &\text{ps: The above equation needs } m\ge (n-l) \text{, so } l\ge (n-m)\\
      =&\left(-i \right)^n \sum_{l=\text{min}\{0,n-m\}}^{n-1} C_n^l m(m-1)\cdots(m+1-n+l)x^{m-n+l} \frac{\partial^l f}{\partial x^l}\\
      =&\sum_{l=\text{min}\{0,n-m\}}^{n-1} \left(-i \right)^{n-l} C_n^l \frac{m!}{(m-n+l)!} x^{m-n+l} p^l\\
  \end{aligned}\nonumber  \\
  \Rightarrow& [p^n,x^m] = \sum_{l=\text{min}\{0,n-m\}}^{n-1} \left(-i \right)^{n-l} \frac{n!m!}{l!(n-l)!(m-n+l)!} x^{m-n+l} p^l   \label{pnxm}
\end{align}
And then, some useful the commutator relations are shown.
\begin{align}
  &\begin{aligned}
    \left[p^2,x^mp^n\right]=&x^m \left[p^2,p^n\right] + \left[p^2,x^m\right]p^n = \left[p^2,x^m\right]p^n=\left(-i \right)^2 \left[ \frac{\partial^2}{\partial x^2}(x^mf) - x^m \frac{\partial^2 f}{\partial x^2} \right] p^n \\
    =&\left(-i \right)^2 \left[ m(m-1)x^{m-2}f + 2mx^{m-1}\frac{\partial f}{\partial x} \right] p^n
  \end{aligned} \nonumber \\
  \Rightarrow& \left[p^2,x^mp^n\right] = -m(m-1)x^{m-2}p^n - 2i  m  x^{m-1}p^{n+1}  \label{p2xmpn}\\
  & \nonumber \\
  &\begin{aligned}
    \left[x^2,p^n\right]=&i^2 \left[  \frac{\partial^2}{\partial p^2}(p^nf) - p^n\frac{\partial^2 f}{\partial p^2}  \right] = i^2 \left[  n(n-1)p^{n-2}f + 2np^{n-1}\frac{\partial f}{\partial p}  \right]\\
    =& -n(n-1)p^{n-2} + 2in p^{n-1}x = -n(n-1)p^{n-2} - 2in \left( \left[x,p^{n-1}\right] - xp^{n-1} \right)\\
    =&-n(n-1)p^{n-2} - 2in \left[  i \left( \frac{\partial}{\partial p} (p^{n-1}f) - p^{n-1}\frac{\partial f}{\partial p} \right) - xp^{n-1}  \right]\\
    =&-n(n-1)p^{n-2} - 2in \left[  i (n-1)p^{n-2} - xp^{n-1}  \right]
  \end{aligned} \nonumber \\
  \Rightarrow& \left[x^2,p^n\right] = 2inx p^{n-1} + n(n-1)p^{n-2} \label{x2pn}\\
  & \nonumber \\
  &\begin{aligned}
    \left[p^2,e^{mx}\right]=& \left(-i \right)^2 \left[  \frac{\partial^2}{\partial x^2}(e^{mx}f)-e^{mx}\frac{\partial^2 f}{\partial x^2}  \right]
    = \left(-i \right)^2 \left[ m^2e^{mx}f+2me^{mx}\frac{\partial f}{\partial x} \right]
  \end{aligned} \nonumber \\
  \Rightarrow& \left[p^2,e^{mx}\right] = -m^2 e^{mx} - 2i  m e^{mx}p \label{p2emx} \\
  & \nonumber \\
  &\begin{aligned}
    \left[p,f\right]=-i \left[ \frac{\partial}{\partial x}(f\psi) - f \frac{\partial \psi}{\partial x} \right] = -i \frac{\partial f}{\partial x}= -if'
  \end{aligned}\label{pf}
\end{align}
Another important relation equation is
\begin{align}
  &\begin{aligned}
    p^ne^{ax}=& \left(-i \right)^n \frac{\partial^n}{\partial x^n}(e^{ax}f)=\left(-i \right)^n \sum_{m=0}^n C_n^m \frac{\partial^m e^{ax}}{\partial x^m} \frac{\partial^{n-m} f}{\partial x^{n-m}}\\
    =& \sum_{m=0}^n C_n^m \left(-i \right)^m a^m e^{ax} p^{n-m} = e^{ax} \sum_{m=0}^n C_n^m \left(-i  a\right)^m p^{n-m}
  \end{aligned} \nonumber \\
  \Rightarrow& p^ne^{ax}= e^{ax} \left(p-i  a\right)^n \label{pneax}
\end{align}

\subsection{Bootstrap Equation of $V(x)$ about One Variable $x$}\label{xOne}
Considering \eqref{EO}\eqref{p2Vx}, when $\mathcal{O}=x^m$, we can get
\begin{gather}
  \begin{aligned}
    \Braket{\mathcal{O}H}=\Braket{x^m p^2} + \Braket{x^m V(x)} = E \Braket{x^m}
  \end{aligned}\label{Exm}\\
  \nonumber \\
  \begin{aligned}
    \Braket{\left[H,\mathcal{O}\right]}=\Braket{\left[p^2,x^m\right]}+\Braket{\left[V(x),x^m\right]}=\Braket{\left[p^2,x^m\right]}=0
  \end{aligned} \nonumber \\
  \overset{\eqref{p2xmpn}}{\Rightarrow} -m(m-1)\Braket{x^{m-2}} - 2im\Braket{x^{m-1}p}=0 \label{p2Vxxm}
\end{gather}
When $\mathcal{O}=x^mp$, we can get
\begin{gather}
  \begin{aligned}
    \Braket{\left[H,\mathcal{O}\right]}=\Braket{\left[p^2,x^mp\right]}+\Braket{\left[V(x),x^mp\right]}=\Braket{\left[p^2,x^m\right]p}+\Braket{x^m\left[V,p\right]}=0
  \end{aligned} \nonumber \\
  \overset{\eqref{p2xmpn}\eqref{pf}}{\Rightarrow} -m(m-1)\Braket{x^{m-2}p} -2im\Braket{x^{m-1}p^2} +i\Braket{x^mV'}=0 \label{p2Vxxmp}
\end{gather}
From \eqref{Exm}(m=m-1),\eqref{p2Vxxm}(m=m-1),\eqref{p2Vxxmp}, we can get
\begin{gather}
  4mE \Braket{x^{m-1}} - 4m \Braket{x^{m-1} V(x)} + m(m-1)(m-2) \Braket{x^{m-3}} - 2 \Braket{x^m V'(x)} =0
\end{gather}

\subsection{Bootstrap Equations of $V(x)$ about Two Variables $x,p$}\label{xTwo}
When the operator $\mathcal{O}$ is
\begin{align}
  \mathcal{O}=x^mp^n\label{xmpn}
\end{align}
From \eqref{EO}\eqref{p2Vx}\eqref{xmpn}, we can get
\begin{gather}
    \Braket{HO}= E\Braket{O} =\Braket{p^2 x^m p^n}+\Braket{V(x)x^m p^n} =\Braket{\left[p^2,x^m\right]p^n} + \Braket{x^m p^{n+2}} + \Braket{V(x)x^m p^n} \nonumber \\
    \overset{\eqref{p2xmpn}}{\Rightarrow} E\Braket{x^m p^n} = -m(m-1)\Braket{x^{m-2}p^n} - 2im\Braket{x^{m-1}p^{n+1}} + \Braket{x^m p^{n+2}} + \Braket{V(x)x^m p^n} 
\end{gather}
and another is
\begin{gather}
  \Braket{\left[H,\mathcal{O}\right]}=\Braket{\left[p^2,x^mp^n\right]}+\Braket{\left[V(x),x^mp^n\right]}=\Braket{\left[p^2,x^m\right]p^n}+\Braket{x^m \left[V(x),p^n\right]}=0 \nonumber \\
  \overset{\eqref{p2xmpn}}{\Rightarrow} -m(m-1) \Braket{x^{m-2}p^n} - 2im \Braket{x^{m-1}p^{n+1}} + \Braket{x^m \left[V(x),p^n\right]} =0 \label{p2Vxxmpn}
\end{gather}

\subsubsection{ $V(x)=x^2$ }\label{harosc}
From \eqref{x2pn}, we can get
\begin{align}
    \begin{aligned}
        \left[V(x),p^n\right] = \left[x^2,p^n\right]= 2inx p^{n-1} + n(n-1)p^{n-2}
    \end{aligned}\label{x2xmpn}
\end{align}
From \eqref{p2Vxxmpn}\eqref{x2xmpn}, we can get
\begin{align}
  \begin{aligned}
    -m(m-1) \Braket{x^{m-2}p^n} - 2im \Braket{x^{m-1}p^{n+1}} + 2in \Braket{x^{m+1}p^{n-1}} + n(n-1) \Braket{x^m p^{n-2}}=0
  \end{aligned}
\end{align}
\subsubsection{$V(x)=x^2+gx^4$}\label{qnharosc}
From \eqref{x2pn},  we can get
\begin{align}
  &\begin{aligned}
    \left[V(x),p^n\right]=&\left[x^2,p^n\right]+g\left[x^4,p^n\right]=\left[x^2,p^n\right]+gx^2\left[x^2,p^n\right]+g\left[x^2,p^n\right]x^2\\
    =& 2in xp^{n-1} + n(n-1)p^{n-2} + 2ign x^3p^{n-1} + gn(n-1)x^2 p^{n-2} + g\left[x^2,p^n\right]x^2
  \end{aligned} \nonumber \\
  &\begin{aligned}
    \left[x^2,p^n\right]x^2 =& 2inx p^{n-1}x^2 + n(n-1)p^{n-2}x^2\\
    =& 2in x^3p^{n-1} + n(n-1)x^2 p^{n-2} - 2in x\left[x^2,p^{n-1}\right] - n(n-1)\left[x^2,p^{n-2}\right]\\
    =& 2in x^3 p^{n-1} + n(n-1)x^2 p^{n-2} - 2in x\left[ 2i(n-1)xp^{n-2}+ (n-1)(n-2)p^{n-3} \right]\\
    & -n(n-1)\left[ 2i(n-2)xp^{n-3}+ (n-2)(n-3)p^{n-4} \right]\\
    =& 2in x^3 p^{n-1} + 5n(n-1)x^2 p^{n-2} - 4in(n-1)(n-2)x p^{n-3} - n(n-1)(n-2)(n-3)p^{n-4}
  \end{aligned} \nonumber \\
  &\begin{aligned}
    \Rightarrow\left[V(x),p^n\right]=& 2inx p^{n-1} + n(n-1)p^{n-2} + 4ign x^3p^{n-1} + 6gn(n-1) x^2p^{n-2}\\
    & - 4ign(n-1)(n-2)x p^{n-3} - gn(n-1)(n-2)(n-3) p^{n-4}
  \end{aligned}\label{x2gx4xmpn}
\end{align}
From \eqref{p2Vxxmpn}\eqref{x2gx4xmpn}, we can get
\begin{gather}
  \begin{aligned}
    &4ign \Braket{x^{m+3}p^{n-1}} + 6gn(n-1) \Braket{x^{m+2}p^{n-2}} - 4ign(n-1)(n-2) \Braket{x^{m+1}p^{n-3}}\\
    &+ 2in \Braket{x^{m+1}p^{n-1}} + n(n-1) \Braket{x^m p^{n-2}} - gn(n-1)(n-2)(n-3) \Braket{x^mp^{n-4}}\\
    &- 2im \Braket{x^{m-1}p^{n+1}} - m(m-1) \Braket{x^{m-2}p^n} = 0
  \end{aligned}
\end{gather}
\subsubsection{$V=\frac{l(l+1)}{r^2}-\frac{N}{r}$}\label{coulomb}
We need two commutators
\begin{gather}
  \begin{aligned}
    [r^{-1},p^n]=& \left(-i\right)^n \left[ r^{-1}\frac{\partial^n f}{\partial r^n} - \frac{\partial^n}{\partial r^n} (r^{-1}f) \right]
    = -\left(-i\right)^n \sum_{t=0}^{n-1} C_n^t \frac{\partial^t r^{-1}}{\partial r^t} \frac{\partial^{n-t}f}{\partial r^{n-t}}\\
    =& -\sum_{t=0}^{n-1} C_n^t \left(-i\right)^t \left(-1\right)^t t! r^{-1-t}p^{n-t}
    = -\sum_{t=0}^{n-1} i^t \frac{n!}{(n-t)!} r^{-1-t}p^{n-t}
  \end{aligned}\\
  \begin{aligned}
    [r^{-2},p^n]=& \left(-i\right)^n \left[ r^{-2}\frac{\partial^n f}{\partial r^n} - \frac{\partial^n}{\partial r^n} (r^{-2}f) \right]
    = -\left(-i\right)^n \sum_{t=0}^{n-1} C_n^t \frac{\partial^t r^{-2}}{\partial r^t} \frac{\partial^{n-t}f}{\partial r^{n-t}}\\
    =& -\sum_{t=0}^{n-1} C_n^t \left(-i\right)^t \left(-1\right)^t (t+1)! r^{-2-t}p^{n-t}
    = -\sum_{t=0}^{n-1} i^t \frac{n!}{(n-t)!(t+1)} r^{-2-t}p^{n-t}
  \end{aligned}
\end{gather}
From \eqref{p2Vxxmpn}, we can get
\begin{align}
  \begin{aligned}
    & -m(m-1) \Braket{x^{m-2}p^n} - 2im \Braket{x^{m-1}p^{n+1}} + N\sum_{t=0}^{n-1} i^t \frac{n!}{(n-t)!} \Braket{r^{m-1-t}p^{n-t}}\\
    & - l(l+1) \sum_{t=0}^{n-1} i^t \frac{n!}{(n-t)!(t+1)} \Braket{r^{m-2-t}p^{n-t}} =0
  \end{aligned}
\end{align}

\subsection{Bootstrap Equation of $V(e^x)$ about One Variable $e^x$}\label{exOne}
When $ \mathcal{O}=e^{mx} $, we can get
\begin{gather}
  \begin{aligned}
   \Braket{ \left[H,\mathcal{O}\right]}=&\Braket{\left[p^2+V(e^x),e^{mx}\right]}=\Braket{\left[p^2,e^{mx}\right]}+\Braket{\left[V(e^x),e^{mx}\right]}=\Braket{\left[p^2,e^{mx}\right]}=0
  \end{aligned} \nonumber \\
  \overset{\eqref{p2emx}}{\Rightarrow}\Braket{ \left[H,\mathcal{O}\right]}= -m^2 \Braket{e^{mx}} - 2im \Braket{e^{mx}p}=0\label{p2Vexemx}\\
  \nonumber \\
  \begin{aligned}
    \Braket{\mathcal{O}H}=\Braket{e^{mx}p^2}+\Braket{e^{mx}V(e^x)}=E\Braket{e^{mx}}
  \end{aligned}\label{emxp2V}
\end{gather}
When $ \mathcal{O}=e^{mx}p $, we can get
\begin{gather}
  \begin{aligned}
    \Braket{\left[H,\mathcal{O}\right]}=&\Braket{\left[p^2,e^{mx}p\right]}+\Braket{\left[V(e^x),e^{mx}p\right]}=\Braket{\left[p^2,e^{mx}\right]p}+\Braket{e^{mx}\left[V(e^x),p\right]}
  \end{aligned} \nonumber \\
  \overset{\eqref{p2emx}\eqref{pf}}{\Rightarrow} -m^2 \Braket{e^{mx}p} - 2im \Braket{e^{mx}p^2} + i\Braket{e^{mx} \frac{\partial V}{\partial x}}=0  \label{p2Vexemxp}
\end{gather}
From \eqref{p2Vexemx},\eqref{emxp2V},\eqref{p2Vexemxp}, we can get
\begin{align}
  m^3 \Braket{e^{mx}} + 4mE\Braket{e^{mx}} - 4m\Braket{e^{mx}V} - 2\Braket{e^{mx}\frac{\partial V}{\partial x}}=0
\end{align}

\subsection{Bootstrap Equations of $V(e^x)$ about Two Variables $e^x,p$}\label{exTwo}
When $\mathcal{O}$ is 
\begin{gather}
  \mathcal{O}=e^{mx}p^n
\end{gather}
from \eqref{EO}, we can get
\begin{gather}
  \begin{aligned}
    \Braket{H\mathcal{O}}=&E\Braket{\mathcal{O}}=\Braket{p^2 e^{mx}p^n} + \Braket{V(e^x)e^{mx}p^n} \overset{\eqref{pneax}}{=} \Braket{e^{mx}\left(p-im\right)^2 p^n} + \Braket{V(e^x)e^{mx}p^n} 
  \end{aligned} \nonumber \\
  \Rightarrow \Braket{e^{mx}p^{n+2}} - 2im\Braket{e^{mx}p^{n+1}} - m^2\Braket{e^{mx}p^n} + \Braket{V(e^x)e^{mx}p^n} = E\Braket{e^{mx}p^n} \label{Eemxpn}
\end{gather}
\begin{gather}
  \begin{aligned}
    \Braket{\left[H,\mathcal{O}\right]}=&\Braket{\left[p^2,e^{mx}p^n\right]}+\Braket{\left[V(e^x),e^{mx}p^n\right]}
    = \Braket{p^2 e^{mx}p^n}-\Braket{e^{mx}p^np^2}+\Braket{e^{mx}\left[V(e^x),p^n\right]}\\
    \overset{\eqref{pneax}}{=}& \Braket{e^{mx}p^{n+2}} - 2im\Braket{e^{mx}p^{n+1}} - m^2\Braket{e^{mx}p^n} - \Braket{e^{mx}p^{n+2}}+\Braket{e^{mx}\left[V(e^x),p^n\right]} = 0\\
  \end{aligned} \nonumber \\
  \Rightarrow -2im\Braket{e^{mx}p^{n+1}} - m^2\Braket{e^{mx}p^n} + \Braket{e^{mx}V(e^x)p^n} - \Braket{e^{mx}p^nV(e^x)} =0 \label{p2Vemxpn}
\end{gather}
$V(e^x)$ is a polynomial about $e^x$, $V(e^x)=\sum_{i=l_1}^{l_2} a_i e^{ix}$. So we can calculate $\Braket{e^{mx}p^nV(e^x)}$ by \eqref{pneax}.
\subsubsection{$V(e^x)=gcosh(x)=g\frac{e^x+e^{-x}}{2}$}\label{Toda}
From \eqref{pneax},we can get
\begin{gather}
  \begin{aligned}
    \Braket{e^{mx}p^nV(e^x)}=& \frac{g}{2}\Braket{e^{mx}p^n e^x} + \frac{g}{2}\Braket{e^{mx}p^n e^{-x}} = \frac{g}{2}\Braket{e^{mx}e^x\left(p-i\right)^n} + \frac{g}{2}\Braket{e^{mx}e^{-x}\left(p+i\right)^n}\\
    =& \frac{g}{2}\sum_{l=0}^n C_n^l \left(-i\right)^{n-l}\Braket{e^{(m+1)x}p^l} + \frac{g}{2}\sum_{l=0}^n C_n^l i^{n-l}\Braket{e^{(m-1)x}p^l}
  \end{aligned}
\end{gather}

\subsection{Bootstrap Equation of $V(x,e^x)$ about Two Variables $x,e^x$}\label{xexTwo}
When $\mathcal{O}=e^{mx}x^n$, from \eqref{EO}\eqref{p2xmpn}\eqref{pneax}, we can get
\begin{gather}
  \begin{gathered}
    \Braket{\mathcal{O}H}=E\Braket{\mathcal{O}}=\Braket{e^{mx}x^n(p^2+V)}
  \end{gathered} \nonumber \\
  \Rightarrow \Braket{e^{mx}x^np^2} + \Braket{e^{mx}x^n V(x,e^x)} = E\Braket{e^{mx}x^n}\label{Eemxxn}\\
  \nonumber\\
  \begin{aligned}
    \Braket{\left[H,\mathcal{O}\right]}=&\Braket{\left[p^2+V,e^{mx}x^n\right]}=\Braket{\left[p^2,e^{mx}x^n\right]}=\Braket{e^{mx}\left[p^2,x^n\right]}+\Braket{\left[p^2,e^{mx}\right]x^n}\\
    =& -n(n-1) \Braket{e^{mx}x^{n-2}} - 2in \Braket{e^{mx}x^{n-1}p} - m^2 \Braket{e^{mx}x^n} - 2im \Braket{e^{mx}px^n}\\
    =& -n(n-1) \Braket{e^{mx}x^{n-2}} - 2in \Braket{e^{mx}x^{n-1}p} - m^2 \Braket{e^{mx}x^n}\\
     & -2mn \Braket{e^{mx}x^{n-1}} - 2im \Braket{e^{mx}x^np} = 0
  \end{aligned}\nonumber\\
  \Rightarrow -m^2 \Braket{e^{mx}x^n} - 2mn\Braket{e^{mx}x^{n-1}} - n(n-1) \Braket{e^{mx}x^{n-2}} - 2i\left[ n\Braket{e^{mx}x^{n-1}p} + m\Braket{e^{mx}x^np}  \right] =0 \label{p2Vemxxn}
\end{gather}
When $\mathcal{O}=e^{mx}x^np$, we have
\begin{align}
  &\begin{aligned}
    \Braket{\left[H,\mathcal{O}\right]}=&\Braket{\left[p^2+V,e^{mx}x^np\right]}=\Braket{\left[p^2,e^{mx}x^n\right]p}+\Braket{e^{mx}x^n\left[V,p\right]}\\
    \overset{\eqref{pf}}{=}& \Braket{e^{mx}\left[p^2,x^n\right]p} + \Braket{\left[p^2,e^{mx}\right]x^n p}+i \Braket{e^{mx}x^n\frac{\partial V}{\partial x}}\\
    \overset{\eqref{p2xmpn}}{=}& -n(n-1) \Braket{e^{mx}x^{n-2}p} - 2in \Braket{e^{mx}x^{n-1}p}+ i\Braket{e^{mx}x^n\frac{\partial V}{\partial x}}\\
    & -m^2 \Braket{e^{mx}x^np} - 2im \Braket{e^{mx}p x^n p} \\
    =& -n(n-1) \Braket{e^{mx}x^{n-2}p} - 2in \Braket{e^{mx}x^{n-1}p} -m^2 \Braket{e^{mx}x^np} + i\Braket{e^{mx}x^n\frac{\partial V}{\partial x}}\\
    & -2im \left( \Braket{e^{mx}\left[p,x^n\right]p} + \Braket{e^{mx}x^n p^2} \right)\\
    \overset{\eqref{pf}}{=}& -n(n-1) \Braket{e^{mx}x^{n-2}p} - 2in \Braket{e^{mx}x^{n-1}p} -m^2 \Braket{e^{mx}x^np} + i\Braket{e^{mx}x^n\frac{\partial V}{\partial x}}\\
    & -2mn \Braket{e^{mx}x^{n-1}p} - 2im \Braket{e^{mx}x^n p^2} =0
  \end{aligned} \nonumber \\
  &\begin{aligned}
    \Rightarrow & -n \left[ (n-1)\Braket{e^{mx}x^{n-2}p}+m\Braket{e^{mx}x^{n-1}p} \right] - m \left[ n\Braket{e^{mx}x^{n-1}p} + m\Braket{e^{mx}x^np} \right]\\
    & -2i \left[ n\Braket{e^{mx}x^{n-1}p^2} + m\Braket{e^{mx}x^np^2}  \right] + i\Braket{e^{mx}x^n\frac{\partial V}{\partial x}} =0
  \end{aligned}\label{p2Vemxxnp}
\end{align}
From \eqref{Eemxxn}\eqref{p2Vemxxn}\eqref{p2Vemxxnp},we can get
\begin{align}
  \begin{aligned}
    & -(m^3+4mE)\Braket{e^{mx}x^n} - (3m^2n+4nE)\Braket{e^{mx}x^{n-1}} - 3mn(n-1)\Braket{e^{mx}x^{n-2}}\\
    & -n(n-1)(n-2)\Braket{e^{mx}x^{n-3}} + 4n\Braket{e^{mx}x^{n-1}V(x,e^x)} + 4m\Braket{e^{mx}x^nV(x,e^x)} \\
    & +2\Braket{e^{mx}x^n\frac{\partial V}{\partial x}} = 0
  \end{aligned}
\end{align}

\subsection{Bootstrap Equation of $V(\theta)$ about one Variable $e^{im\theta}$}\label{eiOne}
Because $\mathcal{O}$ must be cyclical, we consider
\begin{gather}
  \mathcal{O}=e^{im\theta} \text{ or } e^{im\theta}p_\theta
\end{gather}
When $ \mathcal{O}=e^{im\theta} $, we can get
\begin{gather}
  \begin{aligned}
   \Braket{ \left[H,\mathcal{O}\right]}=&\Braket{\left[p_\theta^2+V(\theta),e^{im\theta}\right]}=\Braket{\left[p_\theta^2,e^{im\theta}\right]}+\Braket{\left[V(\theta),e^{im\theta}\right]}=\Braket{\left[p_\theta^2,e^{im\theta}\right]}=0
  \end{aligned} \nonumber \\
  \overset{\eqref{p2emx}}{\Rightarrow}\Braket{ \left[H,\mathcal{O}\right]}= m^2 \Braket{e^{im\theta}} + 2m\Braket{e^{im\theta}p_\theta}=0\label{p2Veim}\\
  \nonumber \\
  \begin{aligned}
    \Braket{\mathcal{O}H}=\Braket{e^{im\theta}p_\theta^2}+\Braket{e^{im\theta}V(\theta)}=E\Braket{e^{im\theta}}
  \end{aligned}\label{eimp2V}
\end{gather}
When $ \mathcal{O}=e^{im\theta}p_\theta $, we can get
\begin{gather}
  \begin{aligned}
    \Braket{\left[H,\mathcal{O}\right]}=&\Braket{\left[p_\theta^2,e^{im\theta}p_\theta\right]}+\Braket{\left[V(\theta),e^{im\theta}p_\theta\right]}=\Braket{\left[p_\theta^2,e^{im\theta}\right]p_\theta}+\Braket{e^{im\theta}\left[V(\theta),p_\theta\right]}
  \end{aligned} \nonumber \\
  \overset{\eqref{p2emx}\eqref{pf}}{\Rightarrow} m^2\Braket{e^{im\theta}p_\theta} + 2m \Braket{e^{im\theta}p_\theta^2} + i\Braket{e^{im\theta} \frac{\partial V}{\partial \theta}}=0  \label{p2Veimp}
\end{gather}
From \eqref{p2Veim},\eqref{eimp2V},\eqref{p2Veimp}, we can get
\begin{align}
  -m^3 \Braket{e^{im\theta}} + 4mE\Braket{e^{im\theta}} - 4m\Braket{e^{im\theta}V(\theta)} + 2i\Braket{e^{im\theta}\frac{\partial V}{\partial \theta}}=0
\end{align}

\subsection{Bootstrap Equations of $V(\theta)$ about Two Variables $e^{im\theta},p_\theta$}\label{eiTwo}
When $\mathcal{O}$ is 
\begin{gather}
  \mathcal{O}=e^{im\theta}p^n
\end{gather}
from \eqref{EO}, we can get
\begin{gather}
  \begin{aligned}
    \Braket{H\mathcal{O}}=&E\Braket{\mathcal{O}}=\Braket{p_\theta^2 e^{im\theta}p_\theta^n} + \Braket{V(\theta)e^{im\theta}p_\theta^n} \overset{\eqref{pneax}}{=} \Braket{e^{im\theta}\left(p+m\right)^2 p_\theta^n} + \Braket{V(\theta)e^{im\theta}p_\theta^n} 
  \end{aligned} \nonumber \\
  \Rightarrow \Braket{e^{im\theta}p_\theta^{n+2}} + 2m\Braket{e^{im\theta}p_\theta^{n+1}} + m^2\Braket{e^{im\theta}p_\theta^n} + \Braket{V(\theta)e^{im\theta}p_\theta^n} = E\Braket{e^{im\theta}p_\theta^n} \label{Eeimpn}
\end{gather}
\begin{gather}
  \begin{aligned}
    \Braket{\left[H,\mathcal{O}\right]}=&\Braket{\left[p_\theta^2,e^{im\theta}p_\theta^n\right]}+\Braket{\left[V(\theta),e^{im\theta}p_\theta^n\right]}
    = \Braket{p_\theta^2 e^{im\theta}p_\theta^n}-\Braket{e^{im\theta}p_\theta^np_\theta^2}+\Braket{e^{im\theta}\left[V(\theta),p_\theta^n\right]}\\
    \overset{\eqref{pneax}}{=}& \Braket{e^{im\theta}p_\theta^{n+2}} + 2m\Braket{e^{im\theta}p_\theta^{n+1}} + m^2\Braket{e^{im\theta}p_\theta^n} - \Braket{e^{im\theta}p_\theta^{n+2}}+\Braket{e^{im\theta}\left[V(\theta),p_\theta^n\right]} = 0\\
  \end{aligned} \nonumber \\
  \Rightarrow 2m\Braket{e^{im\theta}p_\theta^{n+1}} + m^2\Braket{e^{im\theta}p_\theta^n} + \Braket{e^{im\theta}V(\theta)p_\theta^n} - \Braket{e^{im\theta}p_\theta^nV(\theta)} =0 \label{p2Veimpn}
\end{gather}
Because the potential is cyclical, $V(\theta)$ is a polynomial about $e^{i\theta}$.
\begin{gather}
  V(\theta)=\sum_{l=l_1}^{l_2} a_l e^{il\theta}
\end{gather}
So we can calculate $\Braket{e^{im\theta}p_\theta^nV(\theta)}$ by \eqref{pneax}.
\subsubsection{$V(\theta)=gcos\theta=g\frac{e^{i\theta}+e^{-i\theta}}{2}$}\label{cos}
From \eqref{pneax}, we can get
\begin{gather}
  \begin{aligned}
    \Braket{e^{im\theta}p_\theta^nV(\theta)}=& \frac{g}{2}\Braket{e^{im\theta}p_\theta^n e^{i\theta}} + \frac{g}{2}\Braket{e^{im\theta}p_\theta^n e^{-i\theta}} = \frac{g}{2}\Braket{e^{im\theta}e^{i\theta}\left(p_\theta+1\right)^n} + \frac{g}{2}\Braket{e^{im\theta}e^{-i\theta}\left(p_\theta-1\right)^n}\\
    =& \frac{g}{2}\sum_{l=0}^n C_n^l \Braket{e^{i(m+1)\theta}p_\theta^l} + \frac{g}{2}\sum_{l=0}^n C_n^l \left(-i\right)^{n-l} \Braket{e^{i(m-1)\theta}p_\theta^l}
  \end{aligned}
\end{gather}

\newpage
\bibliographystyle{unsrt}  
\bibliography{References}

\end{document}